\documentclass[12pt]{article}
\usepackage{natbib}
\usepackage{amsmath,graphicx,amssymb}
\bibpunct{(}{)}{;}{a}{}{,}

\newtheorem{thm}{Theorem}[section] 
\newtheorem{lemma}{Lemma}[section] 
\newtheorem{cor}{Corollary}[section]

\newcommand{\eps}{\epsilon}

\newcommand{\qed}{{\unskip\nobreak\hfil\penalty50\hskip2em\vadjust{}
            \nobreak\hfil$\Box$\parfillskip=0pt\finalhyphendemerits=0\par}}
\newcommand{\ignore}[1]{}

\usepackage{color}

\begin{document}

\begin{center}
{\bf \large Identification of Signal, Noise, and Indistinguishable Subsets in High-Dimensional Data Analysis}
\end{center}


\begin{center}{\it X. Jessie Jeng}
\footnote{X. Jessie
Jeng is an Assistant Professor in the Department of Statistics, North Carolina State University, 27695. }
\end{center}


\begin{abstract}
Motivated by applications in high-dimensional data analysis where strong signals often stand out easily and weak ones may be indistinguishable from the noise, we develop a statistical framework to provide a novel categorization of the data into the signal, noise, and indistinguishable subsets.
The three-subset categorization is especially relevant under high-dimensionality as a large proportion of signals can be obscured by the large amount of noise.
Understanding the three-subset phenomenon is important for the researchers in real applications to design efficient follow-up studies.
We develop an efficient data-driven procedure to identify the three subsets.
Theoretical study shows that, under certain conditions, only signals are included in the identified signal subset while the remaining signals are included in the identified indistinguishable subsets with high probability.
Moreover, the proposed procedure adapts to the unknown signal intensity, so that the identified indistinguishable subset shrinks with the true indistinguishable subset when signals become stronger.
The procedure is examined and compared with methods based on FDR control using Monte Carlo simulation. Further, it is applied successfully in a real-data application to identify genomic variants having different signal intensity.

\end{abstract}

\vspace{0.1in} {\bf Keywords}: Two-Level Thresholding; Signal detection; False positive control; False negative control; Multiple testing; Variable screening.

\bigskip



\section{Introduction} \label{sec:intro}

The problem of identifying a small number of signals from a large amount of noise is a central topic in modern statistics due to motivations from a wide spectrum of emerging applications. Examples include the detection of astrophysical sources, surveillance for disease outbreaks, identification of causal genetic markers, etc.
In real applications, it is frequently observed that strong signals can easily stand out, while weak ones are often mixed indistinguishably with the noise.
This phenomenon is especially relevant under high-dimensionality as a large proportion of signals can be obscured by the large amount of noise..

In this paper, we aim to extract valuable information from the data by categorizing the data into the signal, noise, and indistinguishable subsets.
More specifically, we want to identify the signal subset in the data which includes only true signals, the noise subset which includes only noise, and the indistinguishable subset, where signals and noise cannot be separated.  To formulate the problem rigorously, let $S_0$ be the collection of noise in the data, and $S_1$ the collection of true signals. The p-value of the data
\begin{equation} \label{def:model}
P_i \sim U 1_{\{i \in S_0\}} + G 1_{\{i \in S_1\}} , \qquad  i \in \{1, \ldots, n\}, \end{equation}
where $U$ is the uniform distribution on $[0, 1]$ and $G$ is some unknown continuous distribution with $G(t) > U(t)$ for all $t \in (0, 1)$.
The $p$-values are ordered as $P_{(1)} \le P_{(2)} \le \ldots \le P_{(n)}$. Define $d_{*}$ as the separation point between the signal and indistinguishable subsets, and $d_{**}$ the separation point between the indistinguishable and  noise subsets, i.e. $d_* = \min\{i : P_{(i)} \text{ from a noise}\}-1$ and $d_{**} = \max\{i: P_{(i)} \text{ from a signal}\}$.
Our goal is to identify the three subsets by estimating $d_*$ and $d_{**}$.

Understanding the three-subset phenomenon can be important for the researchers in real applications to design appropriate follow-up studies and allocate their resources more efficiently.
For instance, candidates in the signal subset may have priority for more focused study, while those in the noise subset can be removed; and, for candidates in the indistinguishable subset, additional data may be collected to further separate weak signals from the noise (\cite{Conneely2010}, \cite{Spencer2009}, \cite{Suresh2012}, etc.).

The proposed framework of three-subset categorization helps to enrich current studies in multiple testing, which largely focus on the dichotomy of rejecting versus not rejecting null hypotheses.
By controlling false positives, multiple testing procedures identify strong signals with high confidence.
Popular criteria for false positive control include family-wise error (FWER) control (\cite{DSB03}, \cite{DVP04}, etc.) and false discovery rate (FDR) control (Benjamini and Hochberg (1995, 2000)). Recent developments in multiple testing focus on improving the power of FDR procedures and controlling FDR under dependence (\cite{GW04}, \cite{STS04}, \cite{ABDJ06}, \cite{SC07}, \cite{Efron07}, \cite{FHG12}, etc.).
These studies, however, would not provide informative results for the weak signals that are indistinguishable from the noise as these signals cannot be separated by controlling the selection of the noise alone. The higher the dimensionality is, the more indistinguishable signals are, and the less efficient the criterion of false positive control could be. This limitation can hinder meaningful applications of multiple testing procedures in ultra-high dimensional data analysis.

To delineate the indistinguishable and  noise subsets would require an adaptive bound for the range of the weak signals.
As the signals are often very sparse compared to the amount of noise,
it is a challenging task to provide a statistical framework to characterize the weak signals. For instance, power analysis in multiple testing is well known to be difficult due to the limited information about the true signals. Another example is in variable selection, where screening procedures are developed to identify and then remove the noise subset (\cite{FL08}, \cite{HM09},  \cite{FSW09}, \cite{FFS11}, \cite{ZLLZ11}, \cite{LZZ12}, etc.). While significant efficiency has been demonstrated for these methods in handling ultra-high dimensional data, setting a good screening parameter remains a difficult problem as it depends on the proportion and intensity of the non-zero coefficients, which are hard to be inferred from the data. Because of the inherent difficulty of weak signal inference, even though the phenomenon of three subsets has been frequently observed (e.g. \cite{DP08}), no rigorous statistical studies have been developed to explore the properties of the three subsets, neither is an efficient categorization method available up-to-date.

In this paper, we demonstrate the existence of the signal, noise, and indistinguishable subsets in Section \ref{sec:2} and connect the results with some recent developments in exact signal recovery. An efficient data-driven procedure called Two-Level Thresholding (TLT) is proposed in Section \ref{sec:MandT} to identify the three subsets by estimating the separation points $d_*$ and $d_{**}$. $d_*$ is estimated by the first level threshold $\hat d_*$, which strongly controls false positives and only selects strong signals with high probability.
The more challenging part is the construction of $\hat d_{**}$, the second level threshold for the separation point between the indistinguishable and  noise subsets.
We develop a data-driven step-down procedure that traverses the ordered $p$-values until all signals are likely to be included.
We show that, under certain conditions, only signals are included in the identified signal subset while the remaining signals are included in the identified indistinguishable subset with high probability.


Besides controlling false positives and false negatives, the proposed TLT procedure adapts to the intensity of the signal, so that the two thresholding levels move closer to each other as signals become stronger and the indistinguishable subset reduces in size.
In the case when all signals are strong enough to be well-separated from the noise, the two thresholding levels converge to a single point.

The construction of TLT is completely data-driven. No prior information of the data distribution is needed; neither are tuning parameters involved in the algorithm. The computation is very fast with complexity $O(n \log n)$. These properties meet the needs of high-dimensional data applications.

The rest of the paper is organized as follows. We first demonstrates the existence of the three subsets in section \ref{sec:2}. Then we introduce
the construction of the TLT procedure with its theoretical properties for the identification of the three subsets in Section \ref{sec:MandT}. Monte Carlo simulations are presented in Section \ref{sec:simulation} to compare the results of TLT with those of the methods based on FDR control. Real-data results are provided in Section \ref{sec:application}
where we apply our procedure to analyze SNP array data.  We conclude in Section \ref{sec:discussion} with further discussions.
The proofs are relegated to the Appendix.

\section{Existence of The Three Subsets} \label{sec:2}

In this section we first present the sufficient and almost necessary conditions for the existence of the signal, noise, and indistinguishable subsets. The results are connected to the recent developments in exact signal recovery. A simulation example is shown to demonstrate  the relationship between the sizes of the three subsets and the signal intensity.
To allow a succinct theoretical study, we assume, in this section, that the observations are generated independently from a normal mixture, i.e.,
\begin{equation} \label{eq:1}
X_i \sim N(0, 1) 1_{\{i \in S_0\}} + N(\mu, 1) 1_{\{i \in S_1\}}, \qquad i \in \{1, \ldots, n\}.
\end{equation}
The following theorem shows the sufficient and almost necessary conditions for the existence of the three subsets.
\begin{thm} \label{thm:3subsets}
Assume model (\ref{eq:1}). Then, asymptotically, the sufficient and almost necessary condition for the existence of the signal subset is
\begin{equation} \label{eq:mu1}
\mu \ge \sqrt{2(1+\eps)\log |S_0|} - \sqrt{2 \log |S_1|},
\end{equation}
for the existence of the indistinguishable subset is
\begin{equation} \label{eq:mu2}
\mu \le \sqrt{2(1-\eps)\log |S_0|} + \sqrt{2 \log |S_1|},
\end{equation}
and for the existence of the noise subset is
\begin{equation} \label{eq:prop}
\log |S_1| \le (1-\eps) \log |S_0|,
\end{equation}
for any $\eps>0$.
\end{thm}
Theorem \ref{thm:3subsets} implies that (a) all three subsets exist when signals are sparse ($|S_1| = o(n)$) and the signal intensity is between the two bounds in (\ref{eq:mu1}) and (\ref{eq:mu2}); (b) when signal intensity is too small ($\mu< \sqrt{2(1+\eps)\log |S_0|} - \sqrt{2 \log |S_1|}$), no signals stand outside the range of the noise, and only the indistinguishable and noise subsets exist; and (c) when signal intensity is large enough ($\mu> \sqrt{2(1-\eps)\log |S_0|} + \sqrt{2 \log |S_1|}$), all signals are excluded from the range of the noise, and only the signal and noise subsets exist. Moreover, (\ref{eq:mu2}) shows that the higher the dimensionality is, the more likely that the indistinguishable subset exists.

\subsection{Connection to Exact Signal Recovery}

It is interesting to note that the sufficient and almost necessary condition for the existence of the indistinguishable subset is closely related to the condition for exact signal recovery in \cite{JJ2012} and in \cite{XCL2011}. Adopting the similar calibrations:
 \begin{equation} \label{eq:2}
 \pi = |S_1|/ n = n^{-\beta}, \quad 0 < \beta < 1, \qquad \text{and} \qquad \mu = \mu_n = \sqrt{2 r \log n}, \quad r>0,
\end{equation}
we have the following result.
\begin{cor} \label{thm:3subsets}
Assume model (\ref{eq:1}) with calibration (\ref{eq:2}). Then, asymptotically, the noise subset always exists, and the sufficient and almost necessary condition for the existence of the signal subset is
\begin{equation} \label{eq:3}
r > (1-\sqrt{1-\beta})^2,
\end{equation}
and for the indistinguishable subset is
\begin{equation} \label{eq:4}
r < (1 + \sqrt{1-\beta})^2.
\end{equation}
\end{cor}
Note that condition (\ref{eq:4}) delineates the complementary set of the exact recovery region in  \cite{JJ2012}. In other words, only when the indistinguishable subset does not exist is it possible to recover all signals with probability $\approx 1$.
It is also interesting to see that condition (\ref{eq:3}) coincides with the detection boundary for the maximum statistic $M_n = \max_{1 \le i \le n}\{X_i\}$ \citep{DJ04}. This shows that only when the signal subset exists is it possible for the maximum statistic $M_n$ to separate the hypotheses $H_0: X_i \sim N(0, 1), 1\le i\le n$ and $H_a: X_i \sim N(0, 1) 1_{\{i \in S_0\}} + N(\mu, 1) 1_{\{i \in S_1\}}, 1\le i \le n$.

\subsection{A Simulation Example}

\begin{figure}[!h]
\begin{center}
\hspace{-0.5in}
\includegraphics[height=0.35\textheight,width=0.48\textwidth]{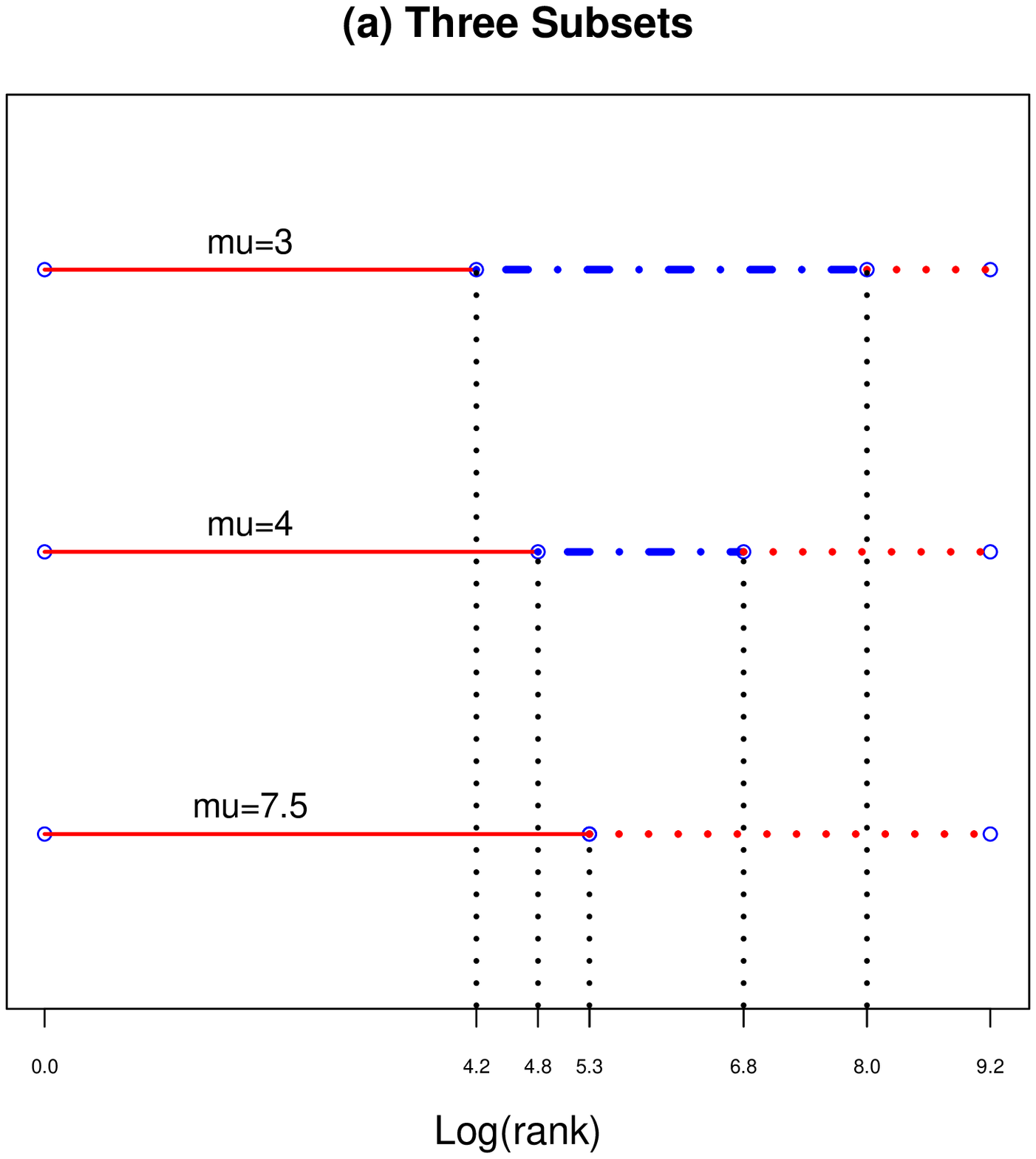}
\hspace{0.15in}
\includegraphics[height=0.35\textheight,width=0.55\textwidth]{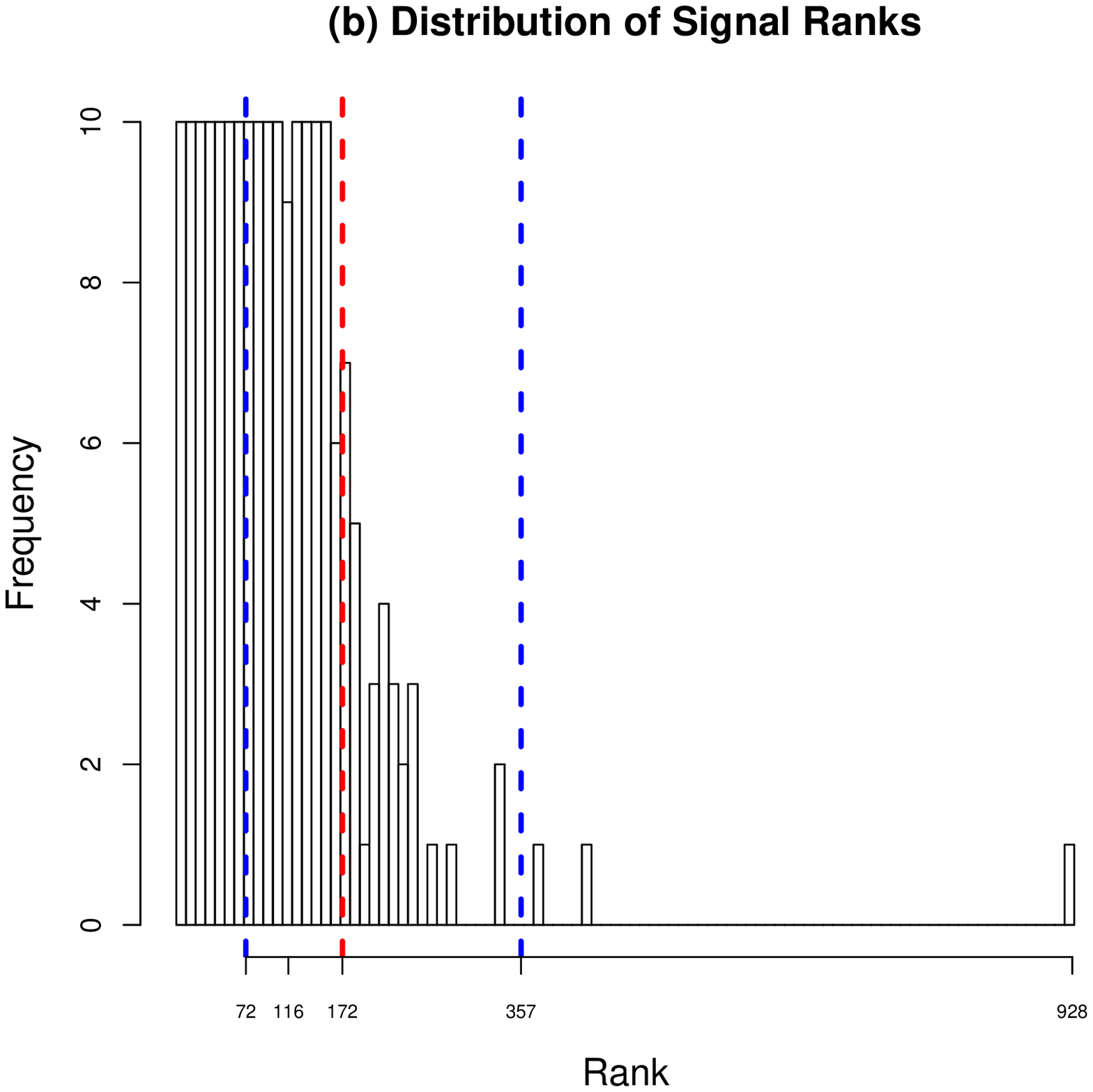}
\end{center}
\vspace{-0.2in}
\caption{\it (a) Three subsets on the rank sequence of ordered $p$-values in log-scale. Signal subset (solid line), indistinguishable subset (dash-and-dot line), and noise subset (dot line) are separated by $d_{*}$ and $d_{**}$. (b) Distribution of the signal ranks when $\mu = 4$. $(d_{*}, ~d_{**})$ are indicated at $(116, ~928)$. Vertical lines at $72$ and $357$ represents the locations of $(\hat d_{*}, ~\hat d_{**})$. The vertical line at $172$ represents the location of the BH-FDR threshold.} \label{Fig:intro}
\end{figure}

The simulation example in this section demonstrates the relationship between the signal intensity and the sizes of the three subsets. The performance of the proposed TLT procedure is also presented in this example. We generated $10,000$ observations and calculate their $p$-values, among which
$2\%$ are from $N(\mu, 1)$ and the rest from $N(0, 1)$. We set $\mu$ at $3$, $4$, and $7.5$. When $\mu=3$,
$(d_*, ~d_{**}) = (65, ~3090)$; when $\mu=4$, $(d_*, ~d_{**}) = (116, ~928)$; and when $\mu=7.5$, $(d_*, ~d_{**}) = (200, ~200)$. The three subsets and $(d_*, ~d_{**})$ are delineated in Figure \ref{Fig:intro} (a) in log-scale for better view. It is clear that, as $\mu$ increases, the signal subset increases to include all true signals, and the indistinguishable subset decreases to an empty set.

For the above example with $\mu=4$ and $(d_*, ~d_{**}) = (116, ~928)$, the distribution of the ranks of the signals is presented in Figure \ref{Fig:intro} (b). Our estimates
 $(\hat d_*, ~\hat d_{**}) = (72, ~357)$, and clearly $\hat d_* < d_*$, so that $p_{(1)}, \ldots, p_{(\hat d_*)}$ are all from signals. $\hat d_{**}=357$, however, is much smaller than $d_{**} = 928$, but $p_{(1)}, \ldots, p_{(\hat d_{**})}$ include $197$ out of $200$ signals, suggesting it as a reasonable estimate for the separation between the indistinguishable and  noise regions. For comparison, the cut-off point of the FDR procedure in \cite{BH95} (BH-FDR) with the control level conventionally set at 0.05 is $172$, which means BH-FDR selects $p_{(1)}, \ldots,
p_{(172)}$ from the ordered $p$-values. The cut-off point of BH-FDR is between $\hat d_*$ and $\hat d_{**}$, and larger than $d_*$. Apparently, BH-FDR selects more signals than $\hat d_*$ and a few noise, but still missing many of the signals.

\section{Identification of The Three Subsets} \label{sec:MandT}

In this section, we first construct the TLT procedure to estimate the separation points between the signal and indistinguishable subsets and between the indistinguishable and  noise subsets, respectively.
Similar to other adaptive procedures in multiple testing, we start with an estimate of the signal proportion $\pi = |S_1|/ (|S_0 \cup S_1|)$. Various estimators have been developed in the literature under certain conditions on the data distribution. For example, \cite{GW04} and \cite{MR06} proposed two proportion estimators under a ``purity" condition on the signal $p$-values. \cite{CJL07}, \cite{JC07}, and \cite{Jin08} developed proportion estimators for normally distributed
observations. Given an estimate $\hat \pi$ for the signal proportion, our estimator for the separation between the signal and indistinguishable subsets is defined as
\begin{equation} \label{def:d*}
 \hat d_*= \max\{i: p_{(i)} < {\alpha_n \over (1-\hat \pi) n }\},
\end{equation}
where $\alpha_n$ is the tolerance level for false positives and $\alpha_n \to 0$ as $n \to \infty$. The choice of the convergence speed of $\alpha_n$ depends on how stringently one wants to control the family-wise type I error. Reasonable choice can be $\alpha_n = 1/\log n$.
$\hat d_*$ can be regarded as an adaptive Bonferonni threshold. Its property of controlling false positives is relatively straightforward. The more challenging part is the construction of $\hat d_{**}$, the estimate of the separation between the indistinguishable and  noise subsets. Even with the help of an estimate for signal proportion, one still does not know where the separation is since the signals are mixed with noise in the indistinguishable subset. Simply cutting at $\hat \pi n$ can include a lot of noise and miss many signals. We propose a data-driven procedure that traverses the ordered $p$-values until all signals are likely to be included. This cut is defined as
\begin{equation} \label{def:d**}
\hat d_{**} = \left\{
\begin{array} {ll}
\hat d_*, &  \hat \pi n \le \hat d_*, \\
\hat \pi n + \min\{j \ge 1: p_{(\hat \pi n + j)} \le F^{-1}_{(j)}(\beta_n) \}, & \mbox{otherwise},
\end{array}
\right.
\end{equation}
where $F^{-1}_{(j)}$ is the inverse cumulative distribution function of Beta$(j, (1-\hat \pi)n-j+1)$, $\beta_n$ is the tolerance level for false negatives, and $\beta_n \to 0$ as $n \to \infty$. A reasonable choice can be $\beta_n = 1 / \log n$.  It is easy to see that $\hat d_{**}$ is always greater than or equal to $\hat d_{_*}$. In the case when $\hat \pi n \le \hat d_*$, all signals are likely to rank before $\hat d_{*}$, then there is no need to go further along the ordered $p$-values, and we set $\hat d_{**} = \hat d_{*}$. On the other hand, $\hat \pi n > \hat d_*$ means that some signals are missing in the first $\hat d_{*}$ ordered $p$-values, so that we need to go further to find all the signals. The search for $\hat d_{**}$ starts at $\hat \pi n$, which is the estimated number of signals, and ends at the smallest $j$ where $p_{(\hat \pi n + j)}$ is no greater than the $\beta_n$-quantile of Beta$(j, (1-\hat \pi)n-j+1)$, which is the distribution of the $j$-th ordered $p$-value from $(1-\hat \pi)n$ noise. The intuition here is that, suppose that not all signals rank before $\hat d_{**}$, then the number of noise in $p_{(1)}, \ldots p_{(\hat d_{**})}$ is likely to be greater than $\hat d_{**} - \hat \pi n$. Denote $\hat j = \hat d_{**} - \hat \pi n$, then the $\hat j$-th ordered $p$-value from $(1-\hat \pi)n$ noise is smaller than $p_{(\hat d_{**})}$. This event, however, has a small probability $\beta_n$ due to the construction of $\hat d_{**}$ where $p_{(\hat d_{**})}\le F^{-1}_{(j)}(\beta_n)$.

Next, we present theoretical results on the properties of the two thresholding levels $\hat d_{*}$ and $\hat d_{**}$.
For simplicity, we utilize the proportion estimator of \cite{MR06}, which is also constructed based on $p$-values.
The estimator, defined as
\begin{equation} \label{def:propEst}
\hat \pi = \max_{1 < i < n/2} {i/n - p_{(i)} - \sqrt{2\log\log n / n} \sqrt{p_{(i)}(1-p_{(i)})} \over 1-p_{(i)}},
\end{equation}
is plugged into (\ref{def:d*}) and (\ref{def:d**}).
Other proportion estimators can be used in the constructions of $\hat d_*$ and $\hat d_{**}$ in a similar way.
The $\hat \pi$ in (\ref{def:propEst}) is a consistent estimator under the following conditions as presented in Theorem 2 and 3 in \cite{MR06}.
Let $\pi = n^{-C}$ for some $C \in [0, 1)$. Assume either
\begin{equation} \label{cond:densePi}
C \in [0, 1/2) \quad \text{and} \quad \inf_{t\in (0,1)} G'(t) = 0,
\end{equation}
or
\begin{equation} \label{cond:sparsePi}
C \in [1/2, 1) \quad \text{and} \quad \text{for all } q \in (0, 1), \quad \lim_{n\to \infty} (\log G^{-1}(q)) / (\log n) = -r, \quad r > 2(C-1/2).  \end{equation}
Condition (\ref{cond:densePi}) considers relatively dense signals with $\pi n \gg \sqrt{n}$; and all we need is the ``purity" condition $\inf_{t\in (0,1)} G'(t) = 0$ (\cite{GW04}, \cite{MR06}). Condition (\ref{cond:sparsePi}) considers sparse signals with $\pi n \le \sqrt{n}$. In this case, stronger condition is needed for signal intensity, which is implied by (\ref{cond:sparsePi}).

Now we show that with high probability, only signals are ranked before $\hat d_*$ and the number of signals ranked before $\hat d_{**}$ converges to $|S_1|$, the total number of signals.
Let
\[
n_0(d) = \text{number of noise in } \{p_{(1)}, \ldots, p_{(d)}\}  \quad \text{and} \quad n_1(d) = \text{number of signals in } \{p_{(1)}, \ldots, p_{(d)}\}
\]
for any integer $d \ge 1$.

\begin{thm} \label{thm:two-cuts}
Assume model (\ref{def:model}) and condition (\ref{cond:densePi}) or (\ref{cond:sparsePi}). Then with high probability, only signals are ranked before the $1^{st}$ thresholding level $\hat d_*$, and the number of signals ranked before the $2^{nd}$ thresholding level $\hat d_{**}$ converges to the total number of signals. That is, as $n \to \infty$,
\begin{equation} \label{0-1}
P(n_0(\hat d_*) > 0) \to 0
\end{equation}
and
\begin{equation} \label{0-2}
P\left({n_1(\hat d_{**}) \over |S_1|} < 1-\eps\right) \to 0
\end{equation}
for any $\eps >0$.
\end{thm}
Theorem \ref{thm:two-cuts} shows that $\hat d_*$ and $\hat d_{**}$ are conservative estimates, which control false positives and false negatives respectively. While one can always achieve conservative estimates at $0$ and $n$, the proposed estimators move closer to each other as signals become stronger and the indistinguishable subset gets smaller.
When all signals are strong enough to be well-separated from the noise, $\hat d_*$ and $\hat d_{**}$
converge to a single point. This adaptivity property of the TLT procedure
 is presented in the following theorem with $\bar G = 1-G$ defined as the survival function of $G$.

\begin{thm} \label{thm:merge}
Assume model (\ref{def:model}). If signals are strong enough, such that $ \pi n \bar G(n^{-r}) \to 0$ for some $r >1$. Then, with high probability, the indistinguishable subset does not exist, and for any $\alpha_n$ satisfying $\log n \ll \log \alpha_n \ll 0$, the signal and noise subsets are consistently separated by $\hat d_* = \hat d_{**}$. That is,
\begin{equation} \label{0-3}
P(\hat d_* = \hat d_{**} = |S_1|) \to 1
\end{equation}
as $n \to \infty$.
\end{thm}
An intuitive understanding for the condition $ \pi n \bar G(n^{-r}) \to 0$, $r >1$, is that $\bar G(n^{-r}) \ll 1/\pi n = 1/n^{1-C} =o(1)$, which means that the total mass of $G$ is asymptotically between 0 and $n^{-r}$. Note that the expectation of the smallest $p$-value from $n$ noise is $n^{-1}$. Therefore, with $r>1$, all the $p$-values of signals are well-separated from all the $p$-values of noise.

Theorem \ref{thm:two-cuts} and \ref{thm:merge} are developed for $\hat d_*$ and $\hat d_{**}$ with $\hat \pi$ defined as in (\ref{def:propEst}). If other proportion estimators are used, conditions in the theorems will be changed accordingly. For example, the proportion estimator in \cite{CJL07} is designed for normally distributed noise and signals.
Utilizing the additional properties of the distribution, this estimator is consistent under a weaker
condition on the signal intensity in the sparse scenario compared to (\ref{cond:sparsePi}) \citep{CJL07}. The theoretical properties of $\hat d_*$ and $\hat d_{**}$ in identifying the signal, noise, and indistinguishable subsets can be proved in a similar way.

In real applications, data may not satisfy the conditions for the existence of a consistent proportion estimator. However, prior knowledge can often allow practitioners to provide a possible range for the signal proportion.  We demonstrate that the study of signal, noise, and indistinguishable subsets can still be carried out utilizing such prior knowledge. Suppose $\pi$ is bounded by
\begin{equation} \label{cond:pibound}
\pi^- \le \pi \le \pi^+.
\end{equation}
Define
\begin{equation} \label{def:tilded*}
 \tilde d_*= \max\{i: p_{(i)} < {\alpha_n \over (1-\pi^-) n }\},
\end{equation}
and
\begin{equation} \label{def:tilded**}
\tilde d_{**} = \left\{
\begin{array} {ll}
\tilde d_*, & \pi^+ n \le \tilde d_*, \\
\pi^+ n + \min\{j \ge 1: p_{(\pi^+ n + j)} \le \tilde F^{-1}_{(j)}(\beta_n) \}, & \mbox{otherwise},
\end{array}
\right.
\end{equation}
where $\tilde F^{-1}_{(j)}$ is the inverse cumulative distribution function for Beta$(j, (1- \pi^-)n-j+1)$. The next theorem states that the modified version $\tilde d_*$ and $\tilde d_{**}$ can still serve as conservative estimates for the separation points $d_*$ and $d_{**}$.

\begin{thm} \label{thm:two-cuts-1}
Assume model (\ref{def:model}) and condition (\ref{cond:pibound}). Then, with high probability,  only signals are ranked before $\tilde d_*$, and the number of signals ranked before $\tilde d_{**}$ converges to the total number of signals.
That is, as $n \to \infty$,
\begin{equation} \label{0-4}
P(n_0(\tilde d_*) > 0)  \to 0
\end{equation}
and
\begin{equation} \label{0-5}
P(n_1(\tilde d_{**}) < |S_1|) \to 0.
\end{equation}
\end{thm}
Although ($\tilde d_*$, $\tilde d_{**}$) may not be as close as ($\hat d_{*}$, $\hat d_{**}$) gets to ($ d_*$, $ d_{**}$),
they can provide useful information of the signal, noise, and indistinguishable subsets in many applications where conditions for the consistency of proportion estimation are hard to be satisfied and some informative prior knowledge of the signal proportion is available.

\section{Simulation} \label{sec:simulation}

In this section, we demonstrate, via simulation studies, the finite sample performance of the TLT procedure on the identification of the signal, noise, and indistinguishable subsets.
In each example, $10,000$ observations are generated, in which the noise data points are sampled from $N(0, \sigma)$ and signals from $N(\mu, 1)$.  The selections of $\hat d_*$ and $\hat d_{**}$ with $\alpha_n = \beta_n = 1/(2 \log n) \approx 0.05$ are compared with those of the BH-FDR with $\alpha=0.05$ \citep{BH95} and the adaptive FDR (\cite{BH00}, \cite{GW04}). Setting $\alpha_n$ at $1/(2 \log n)$ for $n=10,000$ results in a control level close to the conventional level (0.05) used by other methods, so that the results from different methods are comparable. $\beta_n$ is set to be equal to $\alpha_n$ for simplicity.

Among the methods compared, BH-FDR is easiest to implement, while the others require estimating the signal proportion. The estimates $\hat d_*$ and $\hat d_{**}$, the cut-off point of BH-FDR ($t_{FDR}$), the cut-off point of the adaptive FDR ($t_{aFDR}$),
as well as the number of false positives (FP) and the number of false negatives (FN) for each procedure are computed. We repeatedly generate the observations and compute performance measures for 100 times in each simulation example.
The median and mean absolute deviation (MAD) of these measures are reported for more robust comparison results against the outliers in the 100 replications.

\begin{figure}[!h]
\begin{center}
\hspace{-0.5in}
\includegraphics[height=0.2\textheight,width=0.5\textwidth]{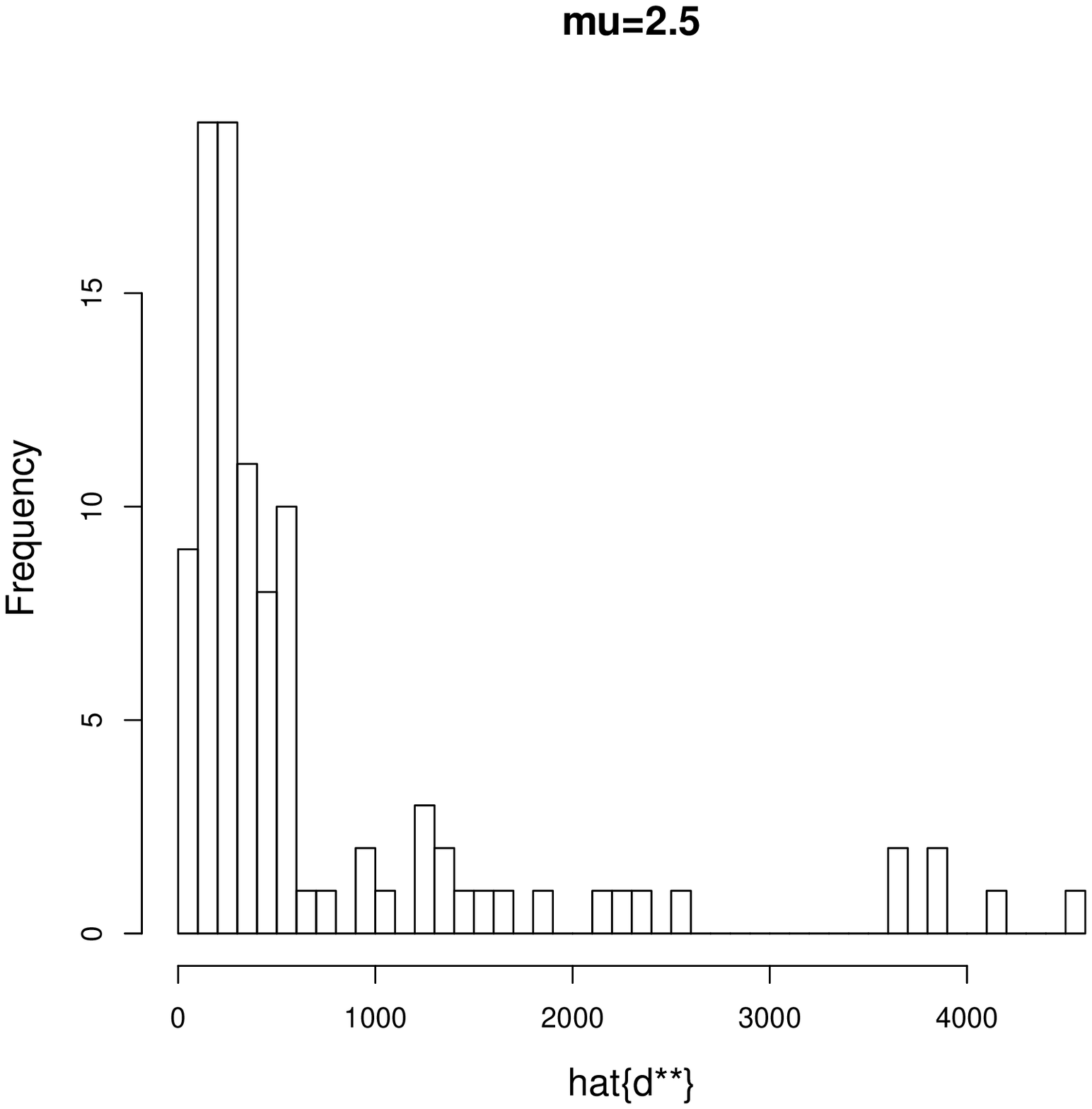}
\hspace{0.15in}
\includegraphics[height=0.2\textheight,width=0.5\textwidth]{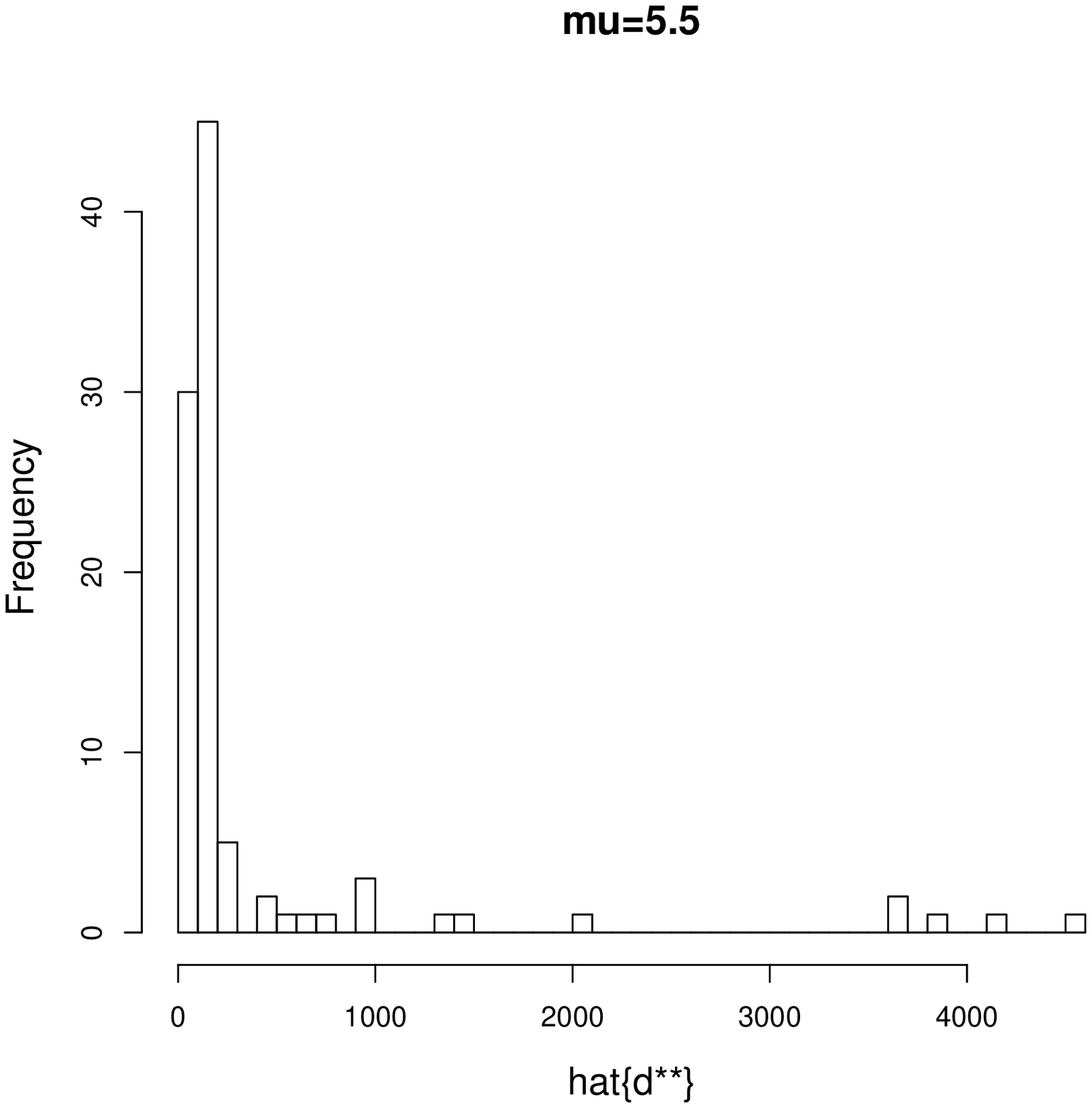}
\end{center}
\vspace{-0.2in}
\caption{\it Histograms of $\hat d_{**}$ for $\mu=2.5$ and $5.5$ from 100 replications. } \label{Fig:Histd}
\end{figure}

Example 1 shows the effect of signal intensity. Set $\sigma = 1$ and $\pi = 0.01$. Signal mean $\mu$ varies from $2.5$ to $5.5$. Since the signal proportion is very small, the results of BH-FDR and the adaptive FDR are very close. To save space, the results of the latter are omitted in this example.
Figure \ref{Fig:Histd} presents the histograms of $\hat d_{**}$ from the 100 replications for $\mu = 2.5$ and $5.5$. It shows that as signal intensity increases, the distribution of $\hat d_{**}$ becomes more concentrated.

Table \ref{table1} shows that the cut-off point of BH-FDR ($t_{FDR}$) is between $\hat d_*$, the estimate of the separation between the signal and indistinguishable subsets (S-I Separation), and $\hat d_{**}$, the estimate of the separation between the indistinguishable and  noise subsets (I-N Separation). As signal intensity increases, the indistinguishable subset shrinks and the cut-off locations of all three procedures move closer. As to the accuracy of identifying the signal, noise, and indistinguishable subsets, it is shown that the FPs of $\hat d_*$ and $t_{FDR}$ are well controlled with $t_{FDR}$ having a bit higher FP when signal intensity increases. This agrees with our intuition since BH-FDR applies a less stringent rule to control false positives.
FP of $\hat d_{**}$ however is not controlled as it is not supposed to be. Interesting results are shown for the FN of $\hat d_{**}$.  Among the $100$ signals, the proportions of mis-specified signals of $\hat d_{**}$ are $28\%$, $11\%$, $3\%$, and $1\%$ for $\mu = 2.5, 3.5, 4.5, 5.5$, respectively.
Compared with the FN of $t_{FDR}$, which has mis-specified proportions of $92\%$, $48\%$, $12\%$, and $1\%$, $\hat d_{**}$ has many fewer false negatives when signals are only moderately strong. This simulation shows that the proposed estimators $\hat{d}_{*}$ and $\hat{d}_{**}$ adapt to the signal intensity, and the identified indistinguishable subset between $\hat{d}_{*}$ and $\hat{d}_{**}$ shrinks with increasing $\mu$.

\begin{table}[!htpb]
\begin{small}
\centering \caption{\it Effect of signal intensity. Median and MAD (in parentheses) of $\hat d_{*}$, $t_{FDR}$, $\hat d_{**}$, and their corresponding FP and FN over  100 replications. $\pi$ is fixed at $1\%$. } \label{table1}
\begin{tabular}{|l|ccc|ccc|ccc|ccc|}
\hline
 & \multicolumn{3}{|c|} {S-I Separation} & \multicolumn{3}{|c|} {BH-FDR} & \multicolumn{3}{|c|} {I-N Separation} \\
 & $\hat d_{*}$ & FP & FN & $t_{FDR}$ & FP & FN & $\hat d_{**}$ & FP & FN \\
\hline
 $\mu=2.5$& 3(1)  &0(0)  &97(1) & 8(4) & 0(0) &92(4) &325(269) & 261(253)& 28(19) \\
$\mu=3.5$ & 17(3) & 0(0) & 83(3) & 54(7) & 2(1) & 48(6) & 194(113) & 103(97) & 11(9) \\
$\mu=4.5$ & 54(4) & 0(0) & 46(5) & 92(4) & 4(3) & 12(3) & 126(44) & 29(34) & 3(3)\\
$\mu=5.5$ & 86(3) & 0(0) & 14(3) & 103(1) & 4(1) & 1(1) & 104(9) & 4(3) & 1(1) \\
\hline
\end{tabular}
\end{small}
\end{table}

\newpage

Example 2 demonstrates the effect of signal proportion. Set $\sigma=1$ and $\mu =3$. The signal proportion $\pi$ changes from $1\%$ to $20\%$. As shown in Table \ref{table2}, when $\pi$ increases, FP of $\hat d_{*}$ remains around $0$. FN of $\hat d_{**}$ is also fairly robust over the different numbers of signals. BH-FDR and adaptive FDR, on the other hand, increases in both FP and FN with increasing signal proportion.

\begin{table}[!htpb]
\begin{scriptsize}
\centering \caption{\it Effect of signal proportion. $\mu$ is fixed at $3$. } \label{table2}
\begin{tabular}{|l|ccc|ccc|ccc|ccc|}
\hline
 & \multicolumn{3}{|c|} {S-I Separation} & \multicolumn{3}{|c|} {BH-FDR}  & \multicolumn{3}{|c|} {adapFDR} & \multicolumn{3}{|c|} {I-N Separation} \\
$|S_1|$ & $\hat d_*$ & FP & FN & $t_{FDR}$ & FP & FN & $t_{aFDR}$ & FP & FN & $\hat d_{**}$ & FP & FN \\
\hline
$100$  & 8(3) & 0(0) & 92(3) & 27(7) & 1(1) & 74(6) & 27(7) & 1(1) & 74(6) & 227(140) & 147(135) & 21(12) \\
$500$  & 40(6) & 0(0) & 460(6) & 255(14) & 11(4) & 255(13) & 259(13) & 12(4) & 253(13) & 1119(494) & 645(462)& 31(28) \\
$1000$ & 81(8.9) & 0(0) & 919(9) & 637(21) & 28(4) & 392(18) & 647(18) & 31(4) & 386(18) & 1960(589) &996(548)& 38(31) \\
$2000$ & 172(11) & 0(0) & 1828(10) & 1485(29) & 59(9) & 575(28) & 1543(32) &72(11) & 529(24) & 3224(660) & 1268(614) & 46(32) \\
\hline
\end{tabular}
\end{scriptsize}
\end{table}

Example 3 has heterogenous noise generated for $10\%$ of the observations. With signal intensity and proportion fixed at $\mu=3.5$ and $\pi=1\%$, the proportion of heterogeneous noise is 10 times the proportion of signals. This example demonstrates a common scenario in real-data applications where unjustified artifacts causes heterogeneity in the background noise.
The heterogeneous noise in this example are randomly generated from $N(0, \sigma)$ with $\sigma \sim$ Gamma$(2, \theta)$. Let the scale parameter $\theta$ vary from $0.5$ to $2$, which results in increasing variability for the noise.
Due to the small signal proportion, the results of the adaptive FDR are very close to those of the BH-FDR and omitted in this example.
Table \ref{table3} shows that FPs of all procedures increase with $\theta$. FNs, on the other hand, are very stable.
Theorem \ref{thm:two-cuts-1} provides some explanation for the robustness of $\hat d_{**}$ in controlling false negatives. Since heterogeneous noise can result in large jumps, the estimated proportion $\hat \pi$ is larger than the true $\pi$. Constructed using this $\hat \pi$, $\hat d_{**}$ is essentially the $\tilde d_{**}$ in (\ref{def:tilded**}), which is built on an upper bound of the true $\pi$. The theoretical property on false negative control is presented in (\ref{0-5}).

\begin{table}[!htpb]
\begin{small}
\centering \caption{\it Robustness for heterogeneous noise. Set $\mu = 3.6$ and $\pi = 1\%$. } \label{table3}
\begin{tabular}{|l|ccc|ccc|ccc|ccc|}
\hline
 & \multicolumn{3}{|c|} {S-I Separation} & \multicolumn{3}{|c|} {BH-FDR} & \multicolumn{3}{|c|} {I-N Separation} \\
 & $\hat d_*$ & FP & FN & $t_{FDR}$ & FP & FN & $\hat d_{**}$ & FP & FN \\
\hline
$\theta = 0.5$ & 22(4) & 5(3) & 82(3) & 69(9) & 15(4) & 45(6) & 196(67) & 107(60) & 12(7) \\
$\theta = 1  $ & 53(7) & 35(6) & 81(4) & 132(12) & 71(9) & 38(4) & 443(180) & 347(174) & 7(4) \\
$\theta = 1.5$ & 94(10) & 75(9) & 80(4) & 195(15) & 130(13) & 35(4) & 556(230) & 459(223) & 7(3) \\
$\theta = 2  $ & 134(12) & 113(10) & 80(4) & 249(12) & 182(10) & 33(4) & 556(179) & 466(175) & 9(4) \\
\hline
\end{tabular}
\end{small}
\end{table}

Example 4 generates autocorrelated observations with $\rho_{ij} = a^{|i-j|}$ for $a = 0, 0.5, 0.7$ and $0.9$. The number of observations are reduced to 1,000 to save computation time. Set $\sigma = 1$, $\pi = 0.05$, $\mu = 3$, and $\alpha_n = \beta_n = 1/\log n$. The results summarized in Table \ref{table5} are quite stable over different values of the autocorrelation parameter $a$ with $\hat d_{**}$ having slightly better control on false negatives for large $a$.

\begin{table}[!htpb]
\begin{small}
\centering \caption{\it Robustness under autocorrelation. Set $\pi = 0.05$ and $\mu = 3$. } \label{table5}
\begin{tabular}{|l|ccc|ccc|ccc|ccc|}
\hline
 & \multicolumn{3}{|c|} {S-I Separation} & \multicolumn{3}{|c|} {BH-FDR} & \multicolumn{3}{|c|} {I-N Separation} \\
 & $\hat d_*$ & FP & FN & $t_{FDR}$ & FP & FN & $\hat d_{**}$ & FP & FN \\
\hline
$a = 0$ & 14(3) & 0(0) & 36(3) & 27(4) & 1(1) & 25(4) & 74(45) & 30(33) & 7(7) \\
$a = 0.5$ & 13(4) & 0(0) & 37(4) & 24(7) & 1(1) & 27(7) & 69(35) & 27(28) & 7(7) \\
$a = 0.7$ & 13(7) & 0(0) & 37(6) & 28(9) & 1(1) & 24(7) & 67(44) & 25(33) & 5(8) \\
$a = 0.9$ & 14(13) & 0(0) & 36(13) & 29(17) & 0(0) & 20(15) & 72(51) & 27(41) & 3(4) \\
\hline
\end{tabular}
\end{small}
\end{table}

\section{Real Application}\label{sec:application}

We apply the three-subset identification to the genotyping data from the Autism Genetics Resource Exchange (AGRE) collection \citep{Bucan09} generated by high-throughput single nucleotide polymorphism (SNP) array technology.
Genotypes in this data set are measured in Log R ratio (LRR), which is calculated at each SNP location as $\log_2(R_{obs}/R_{exp})$, where $R_{obs}$ is the observed total intensity of both major and minor alleles and $R_{exp}$ is computed from a reference genome \citep{Peiffer06}.
LRR data are widely used for detecting copy number variants (CNVs), in which the goal is to identify genomic regions with deletion or duplication of DNA segments \citep{Feuk06}. Such DNA mutations have be reported to play important roles in population diversity and disease association \citep{McCarroll07}.
Due to the fact that the intensity ratio deviates from the baseline in CNV segments, various segment detection methods have been developed to detect CNVs based on SNP array data (\cite{OVLW04}, \cite{ZSJL10}, \cite{SYZ10}, \cite{JCL10}, \cite{JCL12}, etc.)

\begin{figure}[!htpb]
\begin{center}
\includegraphics[height=0.35\textheight,width=0.7\textwidth]{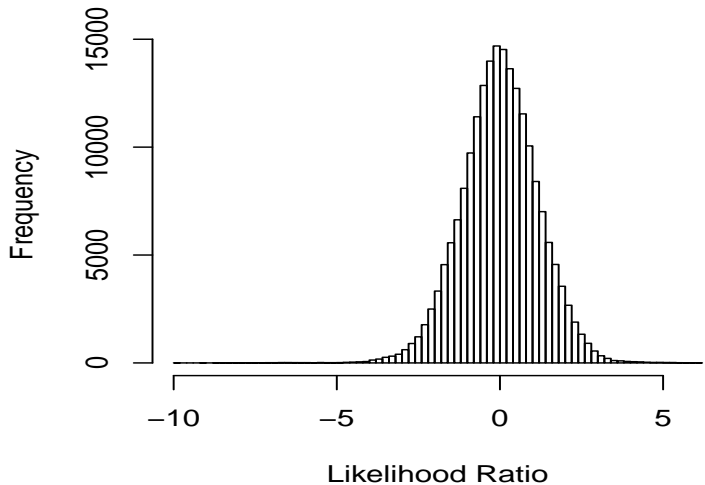}
\end{center}
\vspace{-0.4in}
\caption{\it Histogram of the likelihood ratios of the intervals on Chromosome 19.} \label{Fig:data}
\end{figure}

In this paper, instead of just providing a list of candidates for CNVs, we provide more insight of the data by identifying the signal, noise, and indistinguishable subsets.
We specifically consider the observations on Chromosome 19 for three individuals, which are collected from 9501 SNPs for each individual. The signals are copy number deletions, which may cause LRR to be negative.
For a given individual, LRR observations are first normalized, and then the likelihood ratio is calculated for each interval with length $\le L$ as in \cite{JCL10}. The likelihood ratio of an interval is defined as the standardized sum of observations in that interval, and $L$ is set at 20 as most of the CNVs cover less than 20 SNPs \citep{Zhang09}. There are $n = 9501\times 20 = 190,020$ such likelihood ratio statistics for each individual.
When the distribution of LRR changes in an interval, the corresponding likelihood ratio is expected to deviate from the baseline. Figure \ref{Fig:data} demonstrates the distribution of the likelihood ratios for all the intervals with length $\le L$ on Chromosome 19 of one individual. The outliers in the left tail are likely to come from copy number deletions. The plots are similar for other individuals and are, thus, omitted.

We calculate the $p$-values for these likelihood ratios assuming that the background noise follow $N(0, 1)$ after normalization. The likelihood ratios are locally dependent due to the fact that the intervals are short and overlapping. In this example we treat them as independent observations to illustrate the method. The separations among signal, indistinguishable, and noise subsets are determined by either
$\hat d_{*}$ (\ref{def:d*}) and $\hat d_{**}$ (\ref{def:d**}) or $\tilde d_{*}$ (\ref{def:tilded*}) and $\tilde d_{**}$ (\ref{def:tilded**}). We find that estimating the signal proportion by (\ref{def:propEst}) seems to result in a much larger proportion estimate than commonly expected for SNP array data, possibly due to the artifacts involved in the data generation process \citep{Mar07}.
Thus, we use a more reasonable bound of $0 \le \pi \le 0.005$ for this data set. Setting the upper bound at $0.005$ means that the copy number deletions on Chromosome 19 are approximately less than $50$ \citep{Zhang09}. The signal, noise, and indistinguishable subsets are identified by deriving the cut-off points
 $\tilde d_{*}$ and $\tilde d_{**}$. Because the intervals are overlapping, we only keep intervals having minimum $p$-values among overlapping segments to indicate the locations of copy number deletions. All the other intervals overlapping with them are removed. $\tilde d_{*}$ and $\tilde d_{**}$ are then re-defined as the ranks among these non-overlapping intervals. For the three individuals, ($\tilde d_{*}$, $\tilde d_{**}$) are (2, 18), (1, 76), and (1, 36), respectively.

We further perform validation on the identified signal, noise, and indistinguishable subsets. The candidates in each subset are compared to the reported members in a CNV database maintained in The Centre for Applied Genomics \\ (http://projects.tcag.ca/variation/project.html). A candidate region can overlap with zero, one, or more than one CNVs in the database. The mean value of the number of such CNVs in the database is presented for each subset in Table \ref{table4}. In other words, let $O_j = $ number of CNVs in the database that overlap with the $j$-th candidate in the list of ranked intervals. Define
$\text{ovlap-s} = \text{mean}(O_j , 1\le j \le \tilde d_*), ~ \text{ovlap-i} = \text{mean}(O_j , \tilde d_* < j \le \tilde d_{**}), ~ \text{ovlap-n} = \text{mean}(O_j ,
\tilde d_{**} < j \le \text{ total number of intervals})$.
Table \ref{table4} shows that these mean values, in general, decrease from ovlap-s to ovlap-n. For example, in the identified indistinguishable subset of individual 3, $6.8$ CNVs in the database overlap with each candidate in the identified indistinguishable subset on average, while the number decreases to $2.0$ for the identified noise subset. This agrees with our intuition for the three subsets as larger mean values represents stronger evidence for identifying the true CNVs.
One exception is ovlap-s for individual 3. There is only one candidate in the identified signal subset, which happens to be missed in the database. A possible explanation is that this candidate is a de novo CNV only carried by individual 3.
The sample correlation between the interval length and $O_j$ are $0.17$, $0.28$, and $0.26$ for the three individuals, respectively, indicating that the trend observed in Table \ref{table4} is not likely caused by the length factor.

\begin{table}[!htpb]
\begin{center}
\begin{small}
\caption{\it Estimated separations, $\tilde d_{*}$ and $\tilde d_{**}$, and the mean value of $O_j$ in each subset. } \label{table4}
\begin{tabular}{|l|c||c||c||c||c|}
\hline
 & ovlap-s & $\tilde d_*$ & ovlap-i & $\tilde d_{**}$ & ovlap-n \\
 \hline
Individual 1 & 10.5 & 2& 3.4 & 18& 2.5 \\
Individual 2 & 4 & 1 & 4.7 & 76 & 2.1 \\
Individual 3 & 0 & 1 & 6.8 & 36 & 2.0 \\
\hline
\end{tabular}
\end{small}
\end{center}
\end{table}

\section{Further Discussion} \label{sec:discussion}

In this paper, we developed a novel statistical framework and an efficient TLT procedure to categorize the data into the signal, noise, and indistinguishable subsets. This unique categorization can provide further insight for the data and help the practitioners to design more appropriate follow-up studies to identify the true signals in different subsets.
Another motivation for the new development is its potential to provide an objective criterion for sample-size determination based on the cardinality of the indistinguishable subset.
Unlike traditional sample-size calculation, which is based on a pre-specified level of signal intensity, we may determine whether the sample size is large enough by examining the size of the indistinguishable subset.

Additional insight for the quality of the data may also be achieved by examining the indistinguishable subset. For example, a large indistinguishable subset suggests that there are many small non-null observations, which are either true signals or, very often, caused by artifacts involved during the data generation. Investigating the sources of possible artifacts in follow-up studies may significantly reduce the indistinguishable subset and result in better separation between signals and noise.

We developed two related TLT schemes, one is completely data-driven, the other utilizes prior knowledge on the possible range of the signal proportion. Such flexibility allows practitioners to meet the needs of various applications. The computation for both procedures are very fast.

The study in this paper is based on $p$-values. Other statistics carrying information about signal intensity, such as
the local FDR values (\cite{Efron07}, \cite{SC07}) may be used in place of $p$-values. It will be interesting to investigate this possibility in future research.

In this paper, we assumed independent $p$-values to allow a succinct theoretical study of the new method. Simulation examples in section \ref{sec:simulation} demonstrate the robustness of the proposed method for autocorrelated observations.
We plan to study in depth the three-subset categorization under dependence in future works. We find the recent paper by \cite{FHG12} to be very helpful. According to their work, it is possible to estimate the arbitrary dependence structure of the $p$-values and transform the dependent p-values into weakly dependent ones.

Last but not least, estimating the separation point between the indistinguishable and  noise subsets can be related to the problem of variable screening in high-dimensional regression and can provide new insights on the well-known challenge of screening parameter selection in high-dimensional data analysis.

\section*{Acknowledgement}

We thank Dr. Leonard Stefanski and Dr. John Daye for helpful discussions and comments.

\section*{Appendix: Proofs}

The proofs for theorems in section \ref{sec:2} and \ref{sec:MandT} are provided. A preliminary lemma is first introduced to summarize part of the results in Theorem 1 and 2 in Meinshausen and Rice (2006). The proof of the lemma is omitted.

\begin{lemma} \label{lemma:hatpi}
Assume the same conditions as in Theorem \ref{thm:two-cuts}. Let $\hat \pi$ be defined as in (\ref{def:propEst}). Then for any given $\eps>0$,
\[
P((1-\eps) \pi \le \hat \pi \le \pi) \to 1
\]
as $n \to \infty$.
\end{lemma}

\noindent {\bf Proof of Theorem \ref{thm:3subsets}}

It is sufficient to show the following claims. For any $\eps >0$,
\begin{equation} \label{3-1}
P(\nexists \text{ signal subset}) = o(1) \qquad \text{given} \qquad \mu \ge \sqrt{2(1+\eps)\log |S_0|} - \sqrt{2 \log |S_1|},
\end{equation}
\begin{equation} \label{3-2}
P(\exists \text{ signal subset}) = o(1) \qquad \text{given} \qquad \mu \le \sqrt{2(1-\eps)\log |S_0|} - \sqrt{2 \log |S_1|},
\end{equation}
\begin{equation} \label{3-3}
P(\nexists \text{ indist. subset}) = o(1) \qquad \text{given} \qquad \mu \le \sqrt{2(1-\eps)\log |S_0|} + \sqrt{2 \log |S_1|},
\end{equation}
\begin{equation} \label{3-4}
P(\exists \text{ indist. subset}) = o(1) \qquad \text{given} \qquad \mu \ge \sqrt{2(1+\eps)\log |S_0|} + \sqrt{2 \log |S_1|}.
\end{equation}
\begin{equation} \label{3-0}
P(\nexists \text{ noise subset}) = o(1) \qquad \text{given} \qquad \log |S_1| \le (1-\eps) \log |S_0|,
\end{equation}
\begin{equation} \label{3-00}
P(\exists \text{ noise subset}) = o(1) \qquad \text{given} \qquad \log |S_1| \ge (1+\eps) \log |S_0|.
\end{equation}

Consider (\ref{3-1}) first.
\begin{eqnarray} \label{3-5}
P(\nexists \text{ signal subset}) & = & P(\max\{X_i, i \in S_1\} \le \max\{X_i, i \in S_0  \}) \nonumber\\
 & \le & P(\max\{X_i, i \in S_1\} \le \sqrt{2 \log |S_0|}) + P(\max\{X_i, i \in S_0\} > \sqrt{2 \log |S_0|}) \nonumber\\
 & \le & P(\max\{X_i, i \in S_1\} \le \sqrt{2 \log |S_0|}) + o(1),
\end{eqnarray}
where the last inequality is by the extreme value theory of standard normal random variables. Also,
\begin{eqnarray} \label{3-6}
& & P(\max\{X_i, i \in S_1\} \le \sqrt{2 \log |S_0|}) \nonumber\\
& = & P(\max\{X_i, i \in S_1\} -\mu \le \sqrt{2 \log |S_0|}- \mu )\nonumber \\
& = & P(\max\{X_i, i \in S_1\} -\mu \le \sqrt{2 \log |S_1|} + (\sqrt{2 \log |S_0|}- \mu - \sqrt{2 \log |S_1|})) \nonumber \\
& \le & P(\max\{X_i, i \in S_1\} -\mu \le \sqrt{2 \log |S_1|} + (\sqrt{2 \log |S_0|}- \sqrt{2(1+\eps) \log |S_0|})) \nonumber \\
& = & o(1),
\end{eqnarray}
where the inequality is by $\mu > \sqrt{2(1+\eps)\log |S_0|} - \sqrt{2 \log |S_1|}$. Combining (\ref{3-5}) and (\ref{3-6}) gives (\ref{3-1}).

Next consider (\ref{3-2}).
\begin{eqnarray} \label{3-7}
& & P(\exists \text{ signal subset}) \nonumber \\
& = & P(\max\{X_i, i \in S_1\} > \max\{X_i, i \in S_0  \}) \nonumber\\
 & \le & P(\max\{X_i, i \in S_1\} > \sqrt{2 \log |S_0|} - \log \log n) + P(\max\{X_i, i \in S_0\} < \sqrt{2 \log |S_0|}- \log \log n) \nonumber\\
 & \le & P(\max\{X_i, i \in S_1\} > \sqrt{2 \log |S_0|} - \log \log n) + o(1),
\end{eqnarray}
where the last inequality is by the extreme value theory of standard normal random variables. Also,
\begin{eqnarray} \label{3-8}
& & P(\max\{X_i, i \in S_1\}  > \sqrt{2 \log |S_0|} - \log \log n) \nonumber\\
& = & P(\max\{X_i, i \in S_1\} -\mu > \sqrt{2 \log |S_1|} + (\sqrt{2 \log |S_0|} - \log \log n - \mu - \sqrt{2 \log |S_1|})) \nonumber \\
& \le & P(\max\{X_i, i \in S_1\} -\mu > \sqrt{2 \log |S_1|} + (\sqrt{2 \log |S_0|} - \log \log n - \sqrt{2(1-\eps) \log |S_0|})) \nonumber \\
& = & o(1),
\end{eqnarray}
where the inequality is by $\mu \le \sqrt{2(1-\eps)\log |S_0|} - \sqrt{2 \log |S_1|}$. Combining (\ref{3-7}) and (\ref{3-8}) gives (\ref{3-2}).

The claims in (\ref{3-3}) - (\ref{3-00}) can be proved in similar ways. \\
\qed

\noindent {\bf Proof of Theorem \ref{thm:two-cuts}}

Consider (\ref{0-1}) first. Since
\begin{eqnarray*}
P(n_0(\hat d_*) > 0)
& \le & P(\exists i \in S_0: P_i \le {\alpha_n \over (1-\hat \pi)n}) \\
& \le & (1-\pi) n  \cdot P( P_i \le {\alpha_n \over (1- \pi)n},~ \hat \pi \le \pi) +  P(\hat \pi > \pi)\\
& \le & \alpha_n + o(1),
\end{eqnarray*}
where the third inequality is by Lemma \ref{lemma:hatpi} and the fact that $p$-values from noise are uniformly distributed. Then (\ref{0-1}) follows.

Next, consider (\ref{0-2}). Define $n_1 = |S_1|$. Recall that $\hat j = \hat d_{**} - \hat \pi n$, then
\begin{eqnarray} \label{1-3}
P(n_1(\hat d_{**}) \le (1-\eps) n_1)
& = & P(n_0(\hat d_{**}) > \hat d_{**} - (1-\eps) n_1)  \nonumber \\
& = & P(n_0(\hat d_{**}) >  \hat \pi n +\hat j - (1-\eps) \pi n)  \nonumber \\
& = & P(n_0(\hat d_{**}) > (\hat \pi - (1 - \eps) \pi) n + \hat j ) \nonumber \\
& \le & P(n_0(\hat d_{**}) > (\hat \pi - (1 - \eps) \pi) n + \hat j, ~\hat \pi \ge (1 - \eps) \pi) + P(\hat \pi < (1 - \eps) \pi) \nonumber \\
& \le & P(n_0(\hat d_{**}) > \hat j) + o(1),
\end{eqnarray}
where the first equality is by $\hat d_{**} = n_0(\hat d_{**}) + n_1(\hat d_{**})$, the second equality is by $n_1 = \pi n$, and the last step is by Lemma \ref{lemma:hatpi}.

In the case of $\hat \pi n \le \hat d_{*}$, we have $\hat d_{**} = \hat d_{*} = \hat \pi n$ and $\hat j = 0$. Then
\begin{equation} \label{1-4}
P(n_0(\hat d_{**}) > \hat j)  = P(n_0(\hat d_{*}) > 0)  \to 0
\end{equation}
by (\ref{0-1}). (\ref{0-2}) follows by combining (\ref{1-3}) and (\ref{1-4}).

In the case of $\hat \pi n > \hat d_{*}$, define $P^0_{(j)}$ as the $j$-th smallest $p$-value from $n_0$ noise. Then
\begin{eqnarray} \label{1-5}
P(n_0(\hat d_{**}) > \hat j)
& \le & P(P^0_{(\hat j)} < p_{(\hat d_{**})}) \nonumber \\
& \le & P\left(Beta(\hat j, n_0 - \hat j +1) < F^{-1}_{(\hat j)}(\beta_n)\right) \nonumber \\
& \le & P\left(Beta(\hat j, (1-\hat \pi)n - \hat j +1) < F^{-1}_{(\hat j)}(\beta_n), \hat \pi \le \pi \right) + P(\hat \pi > \pi) \nonumber \\
& = & \beta_n + o(1),
\end{eqnarray}
where the first inequality is because when the elements from $S_0$ are more than $\hat j$ in $\{1, \ldots, \hat d_{**}\}$, the $\hat j$th smallest $p$-value from $S_0$ must rank before the $p$-value at $\hat d_{**}$. The second inequality is by the well-known fact that $P^0_{(j)} \sim Beta(j, n_0-j+1)$ and the construction of $\hat d_{**}$, where
$p_{(\hat d_{**})} \le F^{-1}_{(j)}(\beta_n)$. The last step is by the definition of $F^{-1}_{(j)}$ and Lemma \ref{lemma:hatpi}. Combining (\ref{1-3}) and (\ref{1-5}) gives (\ref{0-2}).
\\

\noindent {\bf Proof of Theorem \ref{thm:merge}}

Defines events $A = \{\hat d_* \ge \hat \pi n \}$, $B = \{\hat d_* \le \pi n\}$, and $C = \{\hat \pi n = \pi n\}$. By the construction of $\hat d_*$ in (\ref{def:d*}), it is enough to show that
\[
P(A \cap B \cap C) \to 1,
\]
which is implied by
\begin{equation} \label{2}
P(A^c) + P(B^c) + P(C^c) \to 0.
\end{equation}

Consider $P(A^c)$ first.
\begin{eqnarray*}
P(A^c) & \le & P(\hat d_* < \pi n) + P(\hat \pi > \pi)\\
& \le & P(\exists i \in S_1: P_i > {\alpha_n \over (1-\hat\pi)n}) + o(1) \\
& \le & \pi n \bar G({\alpha_n \over (1-\hat\pi)n}) +o(1) \\
& \le & \pi n \bar G(n^{-r}) +o(1) = o(1),
\end{eqnarray*}
where the second inequality is by the construction of $\hat d_*$ in (\ref{def:d*}) and Lemma \ref{lemma:hatpi}, the fourth inequality is by ${\alpha_n \over (1-\hat\pi)n} > n^{-r}$ when $\alpha_n \gg n^{-c}$ and $r > 1$, and the last step is by the condition $ \pi n \bar G(n^{-r}) \to 0$.

For $P(B^c)$, it is easy to show that $P(B^c) = P(n_0(\hat d_*) > 0) \to 0$ by similar arguments leading to (\ref{0-1}).

Now consider $P(C^c)$. By lemma \ref{lemma:hatpi}, it is enough to show that
\[
P(\hat \pi n \le \pi n - 1) \to 0,
\]
which is implied by
\begin{equation} \label{2-1}
P({\hat \pi \over \pi} - 1 < - {1\over \pi n}) \to 0.
\end{equation}
Define
\[
F_n(t) = {1\over n} \sum_{i=1}^n 1 (P_i \le t), \qquad U_{n0}(t) = {1\over n_0} \sum_{i=1}^{n_0} 1 (P_i^{(0)} \le t), \qquad G_{n_1}(t) = {1\over n_1} \sum_{i=1}^{n_1} 1 (P_i^{(1)} \le t).
\]
Then, by the construction of $\hat \pi$ in (\ref{def:propEst}), for any $t \in [0,1]$,
\begin{eqnarray*}
{\hat \pi \over \pi}-1 & \ge & {F_n(t) - t - \pi  \over \pi} -{\sqrt{2\log\log n} \sqrt{t(1-t)}  \over \pi\sqrt{n}} \\
& = & {(1-\pi) U_{n0}(t) + \pi G_{n1}(t) - t-\pi \over \pi}  -{\sqrt{2\log\log n} \sqrt{t(1-t)}  \over \pi\sqrt{n}} \\
& = & (G(t) - 1) + (G_{n_1}(t)- G(t)) + {1-\pi \over \pi} (U_{n_0}(t)-t)-t-{\sqrt{2\log\log n} \sqrt{t(1-t)}  \over \pi\sqrt{n}}.
\end{eqnarray*}
Let $t = n^{-r}$. Then by condition $\pi n\bar G(n^{-r}) \to 0$ and $r>1$,
\[
|G(t) - 1)| = \bar G(n^{-r}) = o({1\over \pi n}),
\]
\[
|G_{n_1}(t)- G(t)| = O_p(\sqrt{{G(t)(1-G(t)) \over n_1}}) = O_p(\sqrt{{\bar G(n^{-r}) \over \pi n}}) = o_p({1\over \pi n}),
\]
\[
{1-\pi \over \pi} |U_{n_0}(t)-t| = O_p({1-\pi \over \pi} \sqrt{{t(1-t) \over n_0}}) = O_p({\sqrt{1-\pi} \over \pi} {1\over n^{(1+r)/2}}) = o_p({1\over \pi n}),
\]
\[
{\sqrt{2\log\log n} \sqrt{t(1-t)}  \over \pi\sqrt{n}} = {\sqrt{2\log\log n} \over \pi} {1\over n^{(1+r)/2}} = o({1\over \pi n}).
\]
Therefore, (\ref{2-1}) follows. Combining the above results for $P(A^c)$, $P(B^c)$, and $P(C^c)$ gives (\ref{2}). \\

\noindent {\bf Proof of Theorem \ref{thm:merge}}

The proof of this theorem is similar to that of Theorem \ref{thm:two-cuts} and is, thus, omitted to save space.


\bibliographystyle{asa}
\bibliography{reference}

\end{document}